\documentclass[a4paper,conference]{IEEEtran} 

\IEEEoverridecommandlockouts

\usepackage{cite}
\usepackage{amsmath,amssymb,amsfonts}
\usepackage{algorithmic}
\usepackage{graphicx}
\usepackage{textcomp}
\usepackage{xcolor}
\usepackage{xspace}
\usepackage{tabularx}
\usepackage{booktabs}
\usepackage{hyperref}
\usepackage{balance}
\usepackage{listings}
\usepackage[inline]{enumitem}
\newcommand{\etal}{et al.\xspace}
\newcommand{\ie}{i.e.,\xspace}
\newcommand{\eg}{e.g.,\xspace}
\newcommand{\fig}[1]{Fig.~\ref{#1}}
\newcommand{\lst}[1]{Listing~\ref{#1}}
\newcommand{\tab}[1]{Table~\ref{#1}}

\newcommand{\github}{GitHub\xspace}
\newcommand{\git}{git\xspace}
\newcommand{\python}{Python\xspace}

\newcommand{\changed}[1]{#1}

\usepackage[]{todonotes}

\begin{document}

\title{Evaluating a bot detection model\\
on git commit messages
\thanks{This research is supported by the joint FNRS / FWO Excellence of Science project SECO-ASSIST and FNRS PDR T.0017.18.}
}

\author{
\IEEEauthorblockN{Mehdi Golzadeh}
\IEEEauthorblockA{
\textit{Software Engineering Lab}\\
\textit{University of Mons}, Belgium \\
mehdi.golzadeh@umons.ac.be}
\and
\IEEEauthorblockN{Alexandre Decan}
\IEEEauthorblockA{
\textit{Software Engineering Lab}\\
\textit{University of Mons}, Belgium \\
alexandre.decan@umons.ac.be}
\and
\IEEEauthorblockN{Tom Mens}
\IEEEauthorblockA{
\textit{Software Engineering Lab}\\
\textit{University of Mons}, Belgium \\
tom.mens@umons.ac.be}
}

\maketitle

\IEEEpubid{\begin{minipage}{\textwidth}\ \\[12pt] \centering
    \\
    Copyright 2020 for this paper by its authors. Use permitted under Creative Commons License Attribution 4.0 International (CC BY 4.0).
\end{minipage}}

\begin{abstract}
Detecting the presence of bots in distributed software development activity is very important in order to prevent bias in large-scale socio-technical empirical analyses. In previous work, we proposed a classification model to detect bots in \github repositories based on the pull request and issue comments of \github accounts. The current study generalises the approach to git contributors based on their commit messages. We train and evaluate the classification model on a large dataset of 6,922 git contributors.
The original model based on pull request and issue comments obtained a precision of 0.77 on this dataset. Retraining the classification model on git commit messages increased the precision to 0.80. As a proof-of-concept, we implemented this model in BoDeGiC, an open source command-line tool to detect bots in git repositories.
\end{abstract}


\section{Introduction}
\label{sec:intro}

Many open source software projects rely on development bots or DevBots (hereafter referred to as {\em bots} for short) to reduce the workload of project maintainers and contributors \cite{Erlenhov2019}. These bots play a prominent role in the software development process, especially in online coding platforms. Bots submit pull requests (PR) to update dependencies~\cite{Mirhosseini2017}, merge PRs~\cite{Wessel2018}, submit issues for different problems~\cite{Wessel2018}, help in code review~\cite{Wessel2020ICSME}, and commit code directly to the codebase~\cite{Dey2020MSR}.
\changed{Some studies propose new bots~\cite{Romero2020,James2020botse,Brown2020,Wyrich2019,Matthies2019,Wessel2019}, while other studies aim to help developers analysing the adoption of existing bots~\cite{Wessel2020ICSME,Storey2016,Wessel2020botse}.}

\changed{The presence of bots might introduce bias in large-scale socio-technical empirical studies, especially if these bots are very active. To prevent such bias it is important to distinguish human activity from bot activity in software development projects.
While it is fairly easy for project members to recognise bots in their projects (since they can rely on their project-specific knowledge and experience), this is not necessarily the case for researchers, especially when they are carrying out large-scale ecosystem-level studies involving hundreds of thousands of distinct projects and accounts.}

Previous studies have proposed methods to automatically identify bots by taking into account specific bot characteristics. Dey \etal~\cite{Dey2020MSR} proposed an ensemble technique to identify bots relying on git commit information such as commit messages, contributors name, and commit files.
In a prior study~\cite{Golzadeh2020jss}, we proposed an automatic classification model to identify bots in git repositories hosted in \github based on the content of PR and issue comments.

This paper aims to evaluate the generalisability of the classification model we proposed in~\cite{Golzadeh2020jss} by applying it to git commit messages.
To do so, we evaluate the model on a set of 6,922 contributors extracted from the \changed{dataset provided by Dey \etal}~\cite{Dey2020MSR}. As will be shown, our classification model is able to attain a good precision, even if it was trained on PR and issue comments rather than on git commit messages. Retraining the classifier specifically on git commit messages slightly improves its precision.
To ease the use of our model by practitioners and researchers, we propose \textsf{BoDeGiC}, a command-line tool to detect bots in git repositories.

\section{Background}
\label{sec:background}

In \changed{our} previous work~\cite{Golzadeh2020jss}, we developed and proposed a classification model and associated tool, called \textsf{BoDeGHa}\footnote{\url{https://github.com/mehdigolzadeh/BoDeGHa}} to identify bots in \github repositories based on the comments they made in PRs and issues.
To train and evaluate this model, we created a ground-truth dataset of \github contributors that were manually labeled as bots and humans by three raters.
The evaluation of the model on the test set achieved a very high overall \changed{accuracy}.

In parallel to \changed{our} prior research, Dey \etal~\cite{Dey2020MSR} conducted a similar study aiming to identify bots in git repositories based on git commit data.
They created a ground-truth dataset based on the World of Code (WoC) dataset~\cite{Ma2019} containing 73M \git repositories. Their ground-truth dataset is composed of 13,150 bots and 13,150 humans that were identified using BIN (for Bot Identification by Name), a technique relying on the presence of some keywords (\eg ``bot'') to identify bots.
Based on this ground-truth dataset, they proposed BIMAN (for Bot Identification by commit Message, commit Association, and contributor Name) to detect bots based on commit activit\changed{ies}.
Although both aforementioned studies pursue the same goal, they are different in essence.
The most prominent difference is the type of data on the basis of which bots are being identified.
\changed{PR and issue comments are messages that are being used to explain or discuss issues or PRs. Such comments are not limited in size, in contrast to commit messages that aim to provide one-liners that summarise the changes made in a commit.} 
In~\cite{Golzadeh2020jss} we identified bots based on PR and issue comments, while BIMAN relies on git commit information to \changed{distinguish} bots from humans. As a consequence, the set of contributors considered in both cases is different, since contributors active in PR or issue comments are not necessarily active in code commits, and vice versa.
Another difference is that we restricted our dataset to contributors active in \github while BIMAN works on \changed{all types of git repositories}, even if they are not hosted on \github. Moreover, when a contributor is active in more than one repository, we considered each repository individually while BIMAN aggregates the activity \changed{from} multiple repositories for each contributor. %

\section{Method and approach}
\label{approach}

\subsection{Initial classification model}

In prior work~\cite{Golzadeh2020jss}, we proposed an approach to distinguish bots from humans \changed{based on} their PR and issue commenting activities in \github repositories.  We developed a model (and associated command-line tool) with a high precision to predict whether a contributor is a bot or human based on the comments made in issues and PRs. The underlying idea was that bots perform automated tasks, therefore, they are assumed to have more repetitive comments than humans. To capture this repetition of comments the model was trained using four features related to the comments associated to a contributor:
\begin{enumerate*}[label=(\roman*)]
    \item we measured the number of comment patterns on the basis of a compound comment similarity metric. We hypothesized that the less comment patterns \changed{a contributor has}, the more likely \changed{that} contributor is a bot;
    \item we computed the Gini coefficient to capture the inequality of the number of comments in patterns;
    \item we counted the number of comments since it allows to distinguish between contributors having a similar number of comment patterns;
    \item and we counted the number of empty comments, driven by the assumption that bots are supposed to have meaningful non-empty messages.
\end{enumerate*}
We refer to \cite{Golzadeh2020jss} for the rationale behind \changed{t}hese features and how we computed them.

To evaluate the performance of the model, we relied on a ground-truth dataset composed of 5,000 distinct \github contributors. To create such a ground-truth dataset, we manually labeled each contributor as bot or human with high inter-rater agreement. The final dataset contains 527 bots and 4,473 humans.
We trained and compared various classification models, and achieved \changed{the} highest results ($F1$-score = \changed{0.98}) with a random forest classifier.
Not only does the model perform well in general, it also achieves high precision and recall \changed{for both classes: bots achieved a precision 0.94, a recall of 0.94 and $F1$ of 0.92, and humans achieved a precision of 0.99, recall of 0.99 and $F1$ of 0.99.}
Only \changed{19} out of 211 bots and \changed{13} out of 1,789 humans were misclassified by the model.

%
%
%\subsection{Ground-Truth Commit Message Dataset}
\label{sec:extraction}

Our previous model performed very well to identify bots based on the repetitiveness of their comments, a text-based activity.
Since git commit messages are also text-based, and since we can expect that bots active in commits exhibit a similar kind of \changed{behaviour}, it seems \changed{promising} to apply our model on git commit messages as well.

In this paper, we will first evaluate how our model (trained on PR and issue comments) performs when applied as-is to git commit messages. Then, we will evaluate the approach we developed in~\cite{Golzadeh2020jss} applied on git commit messages, by training a new model.
To do so, we need a labeled dataset of contributors and their commit messages. We rely on the dataset of git commits that was used in~\cite{Dey2020MSR} and has been made publicly available.\footnote{\url{https://zenodo.org/record/4042126}}

We transformed the dataset to conform to the input format required by our classification model.
First of all, as explained in previous section, our model expects \changed{a} set of features \changed{related to comments} of a contributor \changed{in a specific} repository.
Therefore, if a contributor is active in more than one repository, we split its activity by repository.
Our approach being based on the assumption that bots exhibit more repetitive activities, it cannot be applied for contributors that do not have enough activities. In~\cite{Golzadeh2020}, we observed that the performance of the model decreased when the number of comments was below 10.
Similarly, in this work, we will only consider contributors that have at least 10 commit messages.
As in previous work, we rely on 100 commit messages to compute the set of features. %
This upper bound significantly reduces the computational cost, and has been shown to be more than sufficient to achieve a very high precision in~\cite{Golzadeh2020jss}.
Not imposing an upper bound would require to consider all commits for each contributor-repository pair, and some pairs have more than 20,000 commit \changed{messages}. Even for pairs with over 1,000 commit messages the performance begins to slow down considerably.

After having performed these steps, the resulting dataset contains 311,622 commit messages from 6,922 contributors, 3,380 whom have been labeled as bots and 3,542 as humans. This accounts for around 25\% of the original dataset. The dataset characteristics are summarized in \tab{tab:datasetsummary}.

\begin{table}[!tb]
  \centering
  \caption{Summary of the dataset characteristics.}
  \label{tab:datasetsummary}
  \begin{tabular}{l|r}
      \toprule
      \bf original commit dataset from \cite{Dey2020MSR} & \bf number  \\
      \midrule
      \# \git repositories  & 6,394    \\
      \# commits & 311,622  \\
      \# distinct contributors & 6,922 \\
       \quad $\hookrightarrow$ \# bots & 3,380 \\
      \quad $\hookrightarrow$ \# humans & 3,542 \\
      \bottomrule
  \end{tabular}
\end{table}
%
%

%
%
 %
%
%
%
%
%
%

%
%
%

%

%
%
%

%\section{Results}
\label{sec:results}

In this section, we will first evaluate how the classification model trained on PR and issue comments performs when applied as-is to git commit messages. Then, we will evaluate how the approach developed in~\cite{Golzadeh2020jss} applies to git commit messages, by training a new classification model.

We start by applying the \changed{existing} model to see how it performs when applied on a new kind of data (\ie on git commit messages).
For each of the 6,922 contributors in the dataset, we computed the features required by the model, and we asked the model to predict whether the contributor is a bot or a human.
We then compared the prediction with the ground-truth, enabling us to compute the precision $P$ of the model, its recall $R$ and its $F1$-score. The results are reported in \tab{tab:classificationreport}.

\begin{table}[t]
    \caption{Evaluation of the classification model of~\cite{Golzadeh2020jss}.}
    \label{tab:classificationreport}
    \centering
    \begin{tabular}{l|rrrrr}
        {}    &  classified & classified & P  &  R & F1 \\
        & as bot & as human & & & \\
        \midrule
        Bot      &   2,631 (TP) &  749 (FN)&    0.76  &  0.78 &  0.77 \\
        Human      & 848 (FP) &  2,694 (TN)&    0.78  &  0.76  &  0.77 \\
        \midrule
        weighted avg      &   & &    \bf 0.77  &  0.77   &   0.77\\
    \end{tabular}
\end{table}

The model achieved a precision of 0.77, with about 22.1\% (749 out of 3,380) false negatives (\textbf{FN}) of bots misclassified as humans, and about 23.9\% (848 out of 3,542) false positives  (\textbf{FP}) of humans misclassified as bots.
Most contributors are correctly classified as bot or human by the existing model even if it was not trained on git commit messages but on PR and issue comments. \changed{A possible explanation for this result is that even though the model was originally trained on issue and PR comments, it mostly captures the repetitive nature of tasks. Therefore, it shouldn't be that surprising that it also works on commit messages, where we can also expect bots to have repetitive behaviour.}

\medskip

To see how the approach developed in~\cite{Golzadeh2020jss} behaves on git commit messages, we trained a new model on git commit messages as opposed to the previous model that was trained on PR and issue comments.
To do so, we divided the ground-truth dataset into two disjoint subsets. 60\% of the data are used to perform grid-search cross-validation to select the \changed{best} classifier and its parameters. A test set composed of the remaining 40\% is used to evaluate the selected classifier on unseen data.
At the end of the cross-validation set, we obtained 91 different classifiers. %
The performance of these classifiers was measured using traditional performance metrics of precision $P$, recall $R$, and $F1$-score for the population of each class (\ie for bots $B$ and human $H$).
We report the highest $F1$-score for each classifier in \tab{tab:gridsearch}, in descending order.
We retained the random forest (RF) classifier, as it slightly outperforms the other classifiers. Its score was obtained with the \emph{entropy} split criterion, 20 estimators (\ie trees) and a tree depth of 8.

\begin{table*}[!tbph]
    \centering
    \caption{Precision, recall and $F1$ of the best-performing classifiers per classifier family  (in \changed{descending} order of $F1$).}
    \label{tab:gridsearch}
    \begin{tabular}{r|cc|cc|ccc}
        \toprule
        & \multicolumn{2}{c|}{\bf bots} & \multicolumn{2}{c|}{\bf humans} & \multicolumn{3}{c}{\bf overall}\\
        \bf classifier family & $P(B)$ & $R(B)$ & $P(H)$ & $R(H)$ & $P(B\cup H)$ & $R(B\cup H)$  & $F1(B\cup H)$ \\
        \midrule
        random forest ({\bf RF}) &  0.817 &  0.748 &  0.775 &  0.837 &  0.817 &  0.748 &  0.793 \\
        decision trees &  0.845 &  0.698 &  0.750 &  0.876 &  0.845 &  0.698 &  0.787 \\
        support vector machine &  0.798 &  0.735 &  0.762 &  0.819 &  0.798 &  0.735 &  0.777 \\
        logistic regression &  0.807 &  0.720 &  0.755 &  0.832 &  0.807 &  0.720 &  0.776 \\
        k-nearest neighbours &  0.831 &  0.653 &  0.722 &  0.872 &  0.831 &  0.653 &  0.761 \\
        \bottomrule
    \end{tabular}
\end{table*}

We evaluated the selected classifier on the test set containing the remaining 40\% data. This test set includes 2,769 identities, of which 1,417 correspond to humans and 1,352 correspond to bots. The evaluation results are reported in \tab{tab:evalreport}.
With this retrained model about 24.6\% of bots (333 out of 1,352) are misclassified as humans (\textbf{FN}), and about 15.9\% of humans (226 out of 1,417) are misclassified as bots (\textbf{FP}).
Compared to the previous model, the model trained on commit messages \changed{detects humans more accurately}, while the converse can be observed for bots.
\changed{With} a value of $0.80$, the precision of the retrained model is slightly higher than the previous one.

\begin{table}[h]
    \caption{Evaluation of the retrained classification model.}
    \label{tab:evalreport}
    \centering
    \begin{tabular}{l|rrrrr}
        {}    &  classified & classified & P  &  R & F1 \\
        & as bot & as human & & & \\
        \midrule
        Bot      &   1,019 (TP) &  226 (FP) &   0.82  &  0.75 &  0.78 \\
        Human      &  333 (FN)  & 1,191 (TN) &   0.78  &  0.84  &  0.81 \\
        \midrule
        weighted avg      &   & &    \bf 0.80  &  0.80   &   0.80\\
    \end{tabular}
\end{table}\section{A Bot Detector Tool for git repositories}
\label{sec:tool}

In order to enable practitioners to use the classification model, we implemented it through \textsf{BoDeGiC} (Bot Detector for Git Commits)\footnote{\url{https://github.com/mehdigolzadeh/BoDeGiC}}, a command-line tool to detect bots in given git repositories.
The tool analyses the commit messages of each contributor in the specified git repositories and predicts whether the contributor is a bot or a human.
\textsf{BoDeGiC} is implemented in \python 3.7 and is easily installable through \textsf{pip}, the official package manager for \python.
\textsf{BoDeGiC} works in three steps. 
The first step consists of extracting all commit information from the specified \git repository using \texttt{git log}. This step results in a list of commits, authors and their corresponding commit messages.
The second step consists of computing the features to feed the classification model. Features consists of the total number of messages, the number of empty messages, the number of message patterns and the inequality between the number of messages within patterns.
In the third step, we apply the classifier that was trained on \changed{commit messages to the extracted data}. \changed{The tool outputs the prediction made by the model for each contributor.}

\lstset{basicstyle=\ttfamily\footnotesize}

\begin{lstlisting}[caption={List of command-line arguments for \textsf{BoDeGiC} 0.2.0.}\label{fig:command}]
$ bodegic -h
usage: bodegic [-h] [--include [NAME [NAME ...]]]
[--start-date START_DATE][--mapping [MAPPING]] 
[--verbose] [--min-commits MIN_COMMITS][--committer] 
[--max-commits MAX_COMMITS][--text | --csv | --json]
[REPOSITORY [REPOSITORY ...]][--only-predicted] 
\end{lstlisting}

The command-line interface of \textsf{BoDeGiC} is summarized in \lst{fig:command}.
The output and the behaviour of \textsf{BoDeGiC} can be adapted by many optional command-line arguments in several different ways.
By default, \textsf{BoDeGiC} relies on the author names in git commits, but the committer names can be used instead by specifying \texttt{--committer}.
The list of names to consider can be explicitly specified with \texttt{--include}.
\textsf{BoDeGiC} also supports identity merging (\ie when a contributor uses multiple names) through the \texttt{--mapping} parameter. This parameter expects a path to a CSV file specifying how to map names to identities. This file can also be used to ignore specific names, by mapping them to the special ``IGNORE'' identity.
Since the model (and by extension, the tool) requires at least 10 commits for a contributor to generate a prediction, contributors that have commits less than this number are predicted as ``Unknown''. These cases can be excluded from the output by means of the \texttt{--only-predicted} parameter.
Additionally, The minimum and maximum number of commits to consider can be changed respectively with \texttt{--min-commits} and \texttt{--max-commits}.
By default, \textsf{BoDeGiC} outputs one line per author with the predicted class. The set of computed features can be included in this output by adding the \texttt{--verbose} parameter.
Finally, the output of \textsf{BoDeGiC} can be exported in text (by default) or in JSON or as a CSV.

\begin{figure}[!bh]
    \centering
    \includegraphics[width=0.9\columnwidth]{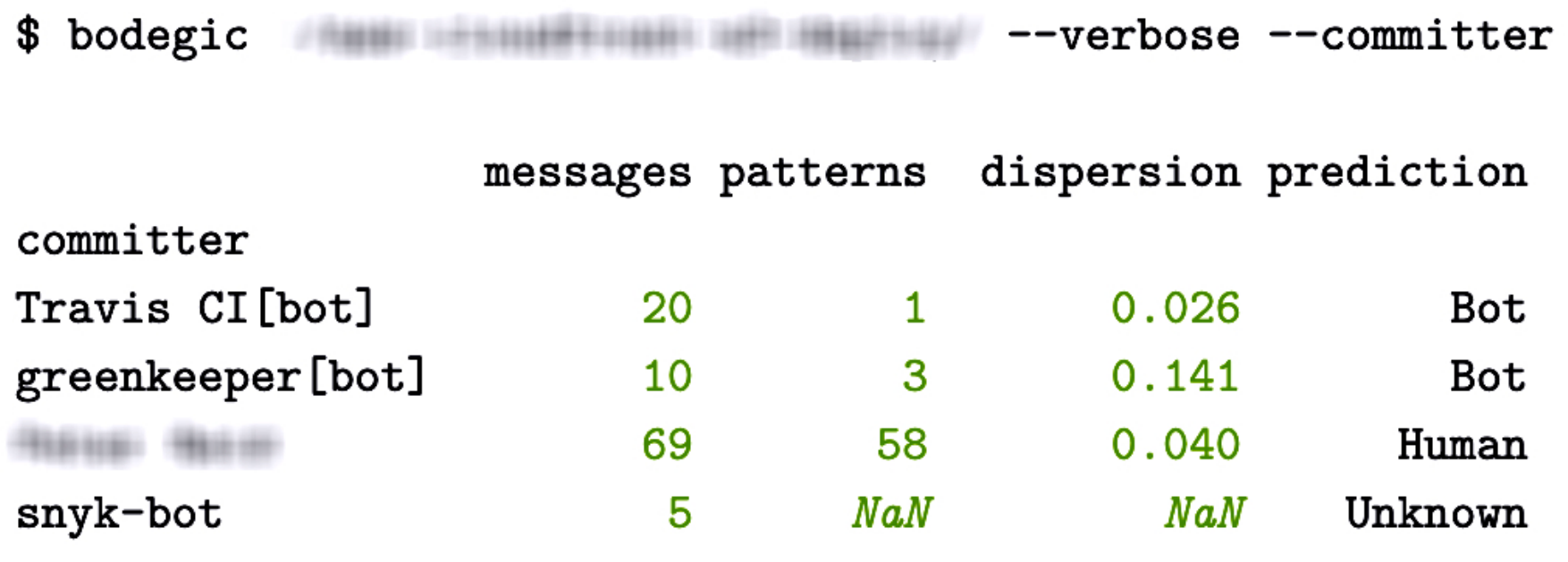}
    \caption{Example of running \textsf{BoDeGiC} (version 0.2.0).}
    \label{fig:output}
\end{figure}

\fig{fig:output} presents the output of \textsf{BoDeGiC} on a randomly chosen \git repository that was analysed on 2020-10-14.
The first columns shows the contributor name, the second column the number of extracted commit messages, the third column the number of computed message patterns, and the fourth column the statistical dispersion of the number of comments per pattern as computed by the Gini inequality index. The last column provides the predicted class of each contributor.

%
%\section{Discussions and Threats}
\label{sec:threats}

The main threats to validity and mitigation strategies mentioned in our prior work~\cite{Golzadeh2020jss} also apply to the current study.

A distinct threat to construct validity stems from the ground-truth dataset that has been used to train and evaluate the models. The dataset was created by other researchers and we have assumed it was correctly built and validated.
Any presence of mislabeled items in that dataset could negatively affect the results of applying our original classification model (\ie that was trained on PR and issue comments) as well as the retrained model that was directly trained on this dataset.

In order to assess to what extent this threat holds, we selected a subset of contributors from the dataset and manually verified whether they are actually humans or bots.
We randomly selected 25 contributors from each category of correctly and incorrectly classified contributors (\ie from TP, TN, FP and FN).
Each of these cases was independently evaluated and labeled by the three authors of this paper.
At the end of this process, we compared the labelings and observed agreement on all 100 cases.
We then compared our own labelings with the actual labels found in the dataset, and observed a disagreement for 19 out of 100 cases. Among these cases, 13 corresponded to bots and 6 to human contributors. The prediction made by our classification model for these 19 cases matched our own labeling, \ie the model was able to correctly predict them.

While manually looking at some other contributors in the dataset, we encountered a few cases we could not agree on because they combine both bot-like and human-like behaviours. We already encountered such ``mixed'' contributors in our previous study ~\cite{Golzadeh2020jss} where we found that some contributors were occasionally relying on tools or bots to automate part of their activit\changed{ies}.
The presence of such mixed cases in git commit messages reinforces our belief that a better definition of ``what a bot is'' is required, with a clearer boundary between humans and bots.
We also believe, in view of these mixed cases, that it might be interesting to identify bots not at the level of a contributor but at the level of its activity. In other words, the question ``Is this contributor a bot?'' would become "Is this contributor activity produced by a bot?''.

\section{Conclusion}
\label{sec:conclusion}

In prior work~\cite{Golzadeh2020jss} we proposed a classification model to predict bots based on features extracted from the PR and issue comments of \github accounts.
The motivation behind the model was that bots tend to carry out repetitive tasks, implying that they are more likely to use repetitive messages.
In this paper, we evaluated to which extent this approach performs on git commit messages, another text-based activity for which we expect bots to exhibit a similar repetitive behaviour.

We first evaluated how the model that we developed and trained on PR and issue comments in \github repositories performs on this new type of data.
We then generalised the classification model by retraining it on a ground-truth dataset composed of commit messages for 6,922 git contributors.
We found that the highest precision was achieved by a random forest classifier, improving the precision of the previous model by 3\% to $0.80$.

We implemented  \textsf{BoDeGiC}, a command-line tool to enable practitioners to take advantage of this classification model.
It analyzes the commit messages of a git repository and predicts whether the contributors are bots or humans.

Although promising, the obtained results suggest that, to achieve a precision comparable to the one we obtained for \github PR and issue comments, it will be necessary to use other data and features in complement to git commit messages.
This is reminiscent of the BIMAN approach~\cite{Dey2020MSR} which, in addition to analysing commit messages (the BIM part of BIMAN) combines two other independent models based on the contributor names (BIN) and features related to files and projects associated with the commits (BICA).
Given that our model is able to accurately capture more bots (+8\%) than BIM, it would be interesting to evaluate how the BIMAN approach would improve by substituting BIM by our classification model.

\bibliographystyle{unsrt}
\balance
\bibliography{biblio.bib}

\end{document}